\documentstyle[12pt,epsf,epsfig]{article}   
\renewcommand{\baselinestretch}{1.5}

 1

\catcode`\@=11 
\makeatletter
\def\@seccntformat#1{\csname the#1\endcsname.\hskip 1em}

\makeatother

\catcode`@=11
\def\chkspace{%
  \relax   
  \begingroup\ifhmode\aftergroup\dochksp@ce\fi\endgroup}
\def\dochksp@ce{%
  \unskip              
  \futurelet\chkspct@k\d@chkspc  
}
\def\d@chkspc{%
  \let\nxtsp@ce=\relax
  \ifx\chkspct@k.\else     
    \ifx\chkspct@k,\else
      \ifx\chkspct@k;\else
        \ifx\chkspct@k!\else
          \ifx\chkspct@k?\else
            \ifx\chkspct@k:\else
              \ifx\chkspct@k)\else
              \ifx\chkspct@k(\else
                \ifx\chkspct@k]\else
                  \ifx\chkspct@k-\else
                    \ifx\chkspct@k\egroup\else  
                      \let\nxtsp@ce=\put@space  
                    \fi
                  \fi
                \fi
              \fi
              \fi
            \fi
          \fi
        \fi
      \fi
    \fi
  \fi
  \nxtsp@ce
}
\def\put@space{$\;$}
\catcode`@=12

\def\gluino{\relax\ifmmode \tilde{g} \else $\tilde{g}$ \fi\chkspace}

\def\qq{\relax\ifmmode q\overline{q}
\else $q\overline{q}$ \fi\chkspace}

\def\bb{\relax\ifmmode b\bar{b}
       \else $b\bar{b}$ \fi\chkspace}
\def\ccrm{\relax\ifmmode {\rm c}\bar{\rm c}
       \else ${\rm c}\bar{\rm c}$ \fi\chkspace}

\def\tt{\relax\ifmmode {\rm t}\bar{\rm t}
       \else ${\rm t}\bar{\rm t}$ \fi\chkspace}
\def\ss{\relax\ifmmode {\rm s}\bar{\rm s}
       \else ${\rm s}\bar{\rm s}$ \fi\chkspace}
\def\uu{\relax\ifmmode {\rm u}\bar{\rm u}
       \else ${\rm u}\bar{\rm u}$ \fi\chkspace}
\def\dd{\relax\ifmmode {\rm d}\bar{\rm d}
       \else ${\rm d}\bar{\rm d}$ \fi\chkspace}

\def\qqg{\relax\ifmmode q\overline{q}g
\else $q\overline{q}g$ \fi\chkspace}
\def\bbg{\relax\ifmmode b\overline{b}g
\else $b\overline{b}g$ \fi\chkspace}

\def\afb{\relax\ifmmode A_{FB} \else
{{$A_{FB}$}}\fi\chkspace}
\def\afbb{\relax\ifmmode A_{FB}^b \else
{{$A_{FB}^b$}}\fi\chkspace}
\def\pafb{\relax\ifmmode \tilde{A}_{FB} \else
{{$\tilde{A}_{FB}$}}\fi\chkspace}
\def\pafbb{\relax\ifmmode \tilde{A}_{FB}^b \else
{{$\tilde{A}_{FB}^b$}}\fi\chkspace}

\def\pafbzo{\relax\ifmmode \tilde{A}_{FB}|_{O(0)} \else
{{$\tilde{A}_{FB}|_{O(0)}$}}\fi\chkspace}
\def\pafbfo{\relax\ifmmode \tilde{A}_{FB}|_{\oalp} \else
{{$\tilde{A}_{FB}|_{\oalp}$}}\fi\chkspace}
\def\pafbso{\relax\ifmmode \tilde{A}_{FB}|_{\oalpsq} \else
{{$\tilde{A}_{FB}|_{\oalpsq}$}}\fi\chkspace}
\def\pafbto{\relax\ifmmode \tilde{A}_{FB}|_{\oalpc} \else
{{$\tilde{A}_{FB}|_{\oalpc}$}}\fi\chkspace}

\def\pafbbzo{\relax\ifmmode \tilde{A}_{FB}^b|_{O(0)} \else
{{$\tilde{A}_{FB}^b|_{O(0)}$}}\fi\chkspace}
\def\pafbbfo{\relax\ifmmode \tilde{A}_{FB}^b|_{\oalp} \else
{{$\tilde{A}_{FB}^b|_{\oalp}$}}\fi\chkspace}
\def\pafbbso{\relax\ifmmode \tilde{A}_{FB}^b|_{\oalpsq} \else
{{$\tilde{A}_{FB}^b|_{\oalpsq}$}}\fi\chkspace}
\def\pafbbto{\relax\ifmmode \tilde{A}_{FB}^b|_{\oalpc} \else
{{$\tilde{A}_{FB}^b|_{\oalpc}$}}\fi\chkspace}

\def\afbo0{\tilde{A}_{FB}|_{O(0)}}
\def\afbo1{\tilde{A}_{FB}|_{\oalp}}
\def\afbo2{\tilde{A}_{FB}|_{\oalpsq}}
\def\afbo3{\tilde{A}_{FB}|_{\oalpc}}

\def\lam{\relax\ifmmode \Lambda_{\overline{MS}}
       \else {{$\Lambda_{\overline{MS}}$}}\fi\chkspace}
\def\lamuds{\relax\ifmmode \Lambda^{(3)}_{\overline{MS}}
       \else {{$\Lambda^{(3)}_{\overline{MS}}$}}\fi\chkspace}
\def\lamudsc{\relax\ifmmode \Lambda^{(4)}_{\overline{MS}}
       \else $\Lambda^{(4)}_{\overline{MS}}$\fi\chkspace}
\def\lamudscb{\relax\ifmmode \Lambda^{(5)}_{\overline{MS}}
       \else $\Lambda^{(5)}_{\overline{MS}}$\fi\chkspace}

\def\alp{\relax\ifmmode \alpha_s\else $\alpha_s$\fi\chkspace}
\def\alpbar{\relax\ifmmode \bar{\alpha_s}
       \else $\bar{\alpha_s}$\fi\chkspace}
\def\alpmz{\relax\ifmmode \alpha_s(M_Z)\else $\alpha_s(M_Z)$\fi\chkspace}
\def\alpmzsq{\relax\ifmmode \alpha_s(M_Z^2)
       \else $\alpha_s(M_Z^2)$\fi\chkspace}

\def\oalp{\relax\ifmmode O(\alpha_s)\else{{O($\alpha_s$)}}\fi\chkspace}
\def\oalpsq{\relax\ifmmode O(\alpha_s^2)
           \else{{O($\alpha_s^2$)}}\fi\chkspace}
\def\oalpc{\relax\ifmmode O(\alpha_s^3)
           \else{{O($\alpha_s^3$)}}\fi\chkspace}
\def\oalpf{\relax\ifmmode O(\alpha_s^4)
           \else{{O($\alpha_s^4$)}}\fi\chkspace}

\def\rb{\relax\ifmmode R_3^b/R_3^{all}
           \else{{$R_3^b/R_3^{all}$}}\fi\chkspace}
\def\rc{\relax\ifmmode R_3^c/R_3^{all}
           \else{{$R_3^c/R_3^{all}$}}\fi\chkspace}
\def\ruds{\relax\ifmmode R_3^{uds}/R_3^{all}
           \else{{$R_3^{uds}/R_3^{all}$}}\fi\chkspace}
\def\ri{\relax\ifmmode R_3^i/R_3^{all}
           \else{{$R_3^i/R_3^{all}$}}\fi\chkspace}
\def\rj{\relax\ifmmode R_3^j/R_3^{all}
           \else{{$R_3^j/R_3^{all}$}}\fi\chkspace}
\def\alpi{\relax\ifmmode \alpha^i_s/\alpha^{all}_s
           \else{{$\alpha^i_s/\alpha^{all}_s$}}\fi\chkspace}

\def\mbz{\relax\ifmmode m_b(M_Z)
           \else{{$m_b(M_Z)$}}\fi\chkspace}
\def\mbb{\relax\ifmmode m_b(M_b)
           \else{{$m_b(M_b)$}}\fi\chkspace}

\def\z0{{$Z^0$}\chkspace}
\def\Dst{\relax\ifmmode {\rm D}^* \else {D$^*$}\fi\chkspace}
\def\Dpl{\relax\ifmmode {\rm D}^+ \else {D$^+$}\fi\chkspace}
\def\D0{\relax\ifmmode {\rm D}^0 \else {D$^0$}\fi\chkspace}
\def\Kst{\relax\ifmmode {\rm K}^* \else {K$^*$}\fi\chkspace}
\def\K0{\relax\ifmmode {\rm K}^0_s \else {K$^0_s$}\fi\chkspace}
\def\Kpl{\relax\ifmmode {\rm K}^+ \else {K$^+$}\fi\chkspace}
\def\Kstz{\relax\ifmmode {\rm K}^{*0} \else {K$^{*0}$}\fi\chkspace}

\def\beq{\begin{equation}}
\def\eeq{\end{equation}}
\def\bea{\begin{eqnarray}}
\def\eea{\end{eqnarray}}

\begin{document}
\thispagestyle{empty}
\begin{flushright}
{\footnotesize\renewcommand{\baselinestretch}{.75}
  SLAC--PUB--8154\\
June 1999\\
}
\end{flushright}

\vskip 1truecm
 
\begin{center}
{\large\bf A Preliminary Direct Measurement of the Parity-Violating 
           Coupling of the Z$^{0}$ to Strange Quarks, $A_s$$^*$}
\end{center}
 
 
\begin{center}
 {\bf The SLD Collaboration$^{**}$}\\
Stanford Linear Accelerator Center \\
Stanford University, Stanford, CA~94309
\end{center}
 
\vspace{0.3cm}
 
\begin{center}
{\bf ABSTRACT}
\end{center}

\noindent
We present an updated direct measurement of the parity-violating 
coupling of the $Z^0$ to strange quarks, $A_s$, derived from the
full SLD data sample of approximately 550,000 hadronic decays of 
Z$^{0}$ bosons produced with a polarized electron beam and recorded 
by the SLD experiment at SLAC between 1993 and 1998. 
$Z^0 \rightarrow s\bar{s}$ events are tagged by the presence in 
each event hemisphere of a high-momentum $K^\pm$, $K_s$ or 
$\Lambda^0$/$\bar\Lambda^0$ identified using the Cherenkov Ring 
Imaging Detector and/or a mass tag. The CCD vertex detector is used 
to suppress the background from heavy-flavor events. The strangeness 
of the tagged particle is used to sign the event thrust axis in the 
direction of the initial $s$ quark. The coupling $A_s$ is obtained 
directly from a measurement of the left-right-forward-backward 
production asymmetry in polar angle of the tagged $s$ quark.
The background from $u\bar{u}$ and $d\bar{d}$ events is measured 
from the data, as is the analyzing power of the method for $s\bar{s}$ 
events. We measure:
\begin{eqnarray*}
  A_s = 0.85 \pm 0.06(stat.) \pm 0.07(syst.) (preliminary).
\end{eqnarray*}


\vspace{0.3cm}

\vfill
\noindent
Contributed to the International Europhysics Conference on High Energy Physics,
15-21 July 1999, Tampere, Finland; Ref. 6\_165, and to the XIXth International 
Symposium on Lepton and Photon Interactions, August 9-14 1999, Stanford, USA.

{\footnotesize
$^*$ Work supported by Department of Energy contract DE-AC03-76SF00515 (SLAC).}

\eject

\section{Introduction}

Measurements of the fermion production asymmetries in the process
$e^+e^- \rightarrow Z^0 \rightarrow f\bar{f}$ provide information on the 
extent of parity violation in the coupling of the $Z^0$ boson to fermions 
of type $f$.
At Born level, the differential production cross section can be expressed 
in terms of $x=\cos\theta$, where $\theta$ is the polar angle of the 
final state fermion $f$ with respect to the electron beam direction:
\begin{equation}
     \sigma^f(x) =     \frac{d\sigma^f}{dx}
            \propto (1-A_e P_e)(1 + x^2) + 2A_f(A_e-P_e)x ,
\end{equation}
where $P_e$ is the longitudinal polarization of the electron beam, the 
positron beam is assumed unpolarized, and the coupling parameters 
$A_f=2v_f a_f/(v_f^2 + a_f^2)$ are defined in terms of the vector ($v_f$) 
and axial-vector ($a_f$) couplings of the $Z^0$ to fermion $f$.
The Standard Model (SM) predictions for the values of the coupling 
parameters, assuming $\sin^2\theta_W=0.23$, are 
$A_e = A_{\mu} = A_{\tau} \simeq 0.16$, $A_u = A_c = A_t \simeq 0.67$ 
and $A_d = A_s = A_b \simeq 0.94$.

If one measures the polar angle distribution for a given final state $
f\bar{f}$, one can derive the forward-backward production asymmetry:
\begin{equation}
 A^f_{FB}(x) = \frac{\sigma^f(x) - \sigma^f(-x)}{\sigma^f(x) + \sigma^f(-x)}   
             = 2 A_f \frac{A_e - P_e}{1 - A_e P_e} \frac{x}{1+x^2}
\end{equation}      
which depends on both the initial and final state coupling parameters as 
well as on the beam polarization.  For zero polarization, one measures 
the product of couplings $A_e A_f$, which has a rather small value 
since $A_e \simeq 0.16$.

If one measures the distributions in equal luminosity samples taken with
negative $(L)$ and positive $(R)$ beam polarization of magnitude $P_e$, 
then one can derive the left-right-forward-backward asymmetry:
\begin{equation} 
\tilde{A}^f_{FB}(x) = \frac{(\sigma^f_L(x) + \sigma^f_R(-x)) - 
                            (\sigma^f_R(x) + \sigma^f_L(-x))}
                           {(\sigma^f_L(x) + \sigma^f_R(-x)) + 
                            (\sigma^f_R(x) + \sigma^f_L(-x))}      
                    = 2 |P_e| A_f \frac{x}{1+x^2}
\end{equation}    
which is insensitive to the initial state coupling.

It is important to measure as many of these coupling parameters as possible, 
in order to test the SM ansatz of lepton, up-quark and down-quark
universality, respectively.
A number of previous measurements have been made by experiments at LEP and SLC
of $A_e, A_{\mu}, A_{\tau}, A_c$ and $A_b$~\cite{ewlepslc}.
The leptonic final states are identified easily by their low track
multiplicities and identification of the stable leptons.
The $c\bar{c}$ final state can be identified by exclusive or partial
reconstruction of the leading charmed hadron in a hadronic jet.
The $b\bar{b}$ final state can be tagged by the presence of a lepton with
high momentum transverse to the jet axis or of a decay vertex displaced from 
the primary interaction point, indicating the presence of a leading $B$ hadron 
in the jet.
In contrast, very few measurements exist for the light flavor quarks, due to 
the difficulty of tagging specific light flavors.
It has recently been demonstrated experimentally~\cite{lpprl} that light
flavored jets can be tagged by the presence of a high-momentum `leading' 
identified particle that has a valence quark of the desired flavor, for 
example a $K^-$ ($K^+$) meson could tag an $s$ ($\bar{s}$) jet.
However the background from other light flavors (a $\bar{u}$ jet can also
produce a leading $K^-$), decays of $B$ and $D$ hadrons, and non-leading kaons
in events of all flavors is large, and neither the signal nor the background
has been well measured experimentally.

The DELPHI collaboration has measured~\cite{delphi} the polar angle 
production asymmetries of $K^\pm$
mesons in the momentum range $10<p<18$ GeV/c, $\Lambda^0$/$\bar\Lambda^0$ 
baryons in the momentum range $11.41<p<22.82$ GeV/c and neutral hadronic 
calorimeter clusters with $E>15$ GeV, from which they have measured 
$A_{FB}^s= 0.131\pm 0.035(stat.)\pm 0.013(syst.)$ and
$A_{FB}^{d,s}=0.112\pm 0.031(stat.)\pm 0.054(syst.)$, respectively,
where $A_{FB}^{d,s}$ denotes a measurement assuming $A_{FB}^{d} = A_{FB}^{s}$.
However the extraction of the coupling parameters from the measured production 
asymmetries is model dependent.
The OPAL collaboration has measured~\cite{opal} the production asymmetries 
of a number of identified particle species with $x_p = 2p /E_{cm} \geq 0.5$, 
where $E_{cm}$ denotes the center-of-mass energy
in the event, and has determined most of the background contributions and 
analyzing powers from double-tagged events in the data.
This eliminates most of the model dependence, but results in limited 
statistical precision, yielding
$A_{FB}^u=0.040\pm 0.067(stat.)\pm 0.028(syst.)$, 
$A_{FB}^{d,s}=0.068\pm 0.035(stat.)\pm 0.011(syst.)$.

In this paper we present a measurement of the coupling parameter for strange
quarks, $A_s$, using the sample of 550,000 hadronic $Z^0$ decays recorded by 
the SLD experiment at the SLAC Linear Collider between 1993 and 1998, with an
average electron beam polarization of 73\%.
Hemispheres are tagged as $s$ ($\bar{s}$) by the presence of a $K^-$ ($K^+$) 
meson identified by the Cherenkov Ring Imaging Detector (CRID) or a 
$\Lambda^0$ ($\bar{\Lambda}^0$) hyperon tagged using a combination of flight 
distance and CRID information.
The background from heavy flavor events ($c\bar{c}$ and $b\bar{b}$) was
suppressed by using B and D hadron lifetime information, 
allowing the use of relatively 
low-momentum identified kaons to tag $s$ or $\bar s$ jets.
The background from the other light flavors ($u\bar{u}$ and $d\bar{d}$) was
suppressed by the additional requirement of a high-momentum identified strange 
particle in the opposite hemisphere of the event.
This analysis is discussed in section~3. The coupling parameter was extracted 
from a simultaneous unbinned maximum likelihood fit to the polar angle
distributions measured with left- and right-handed electron beams, as 
discussed in section 4;
this is a direct measurement, i.e. it is insensitive to the initial state 
coupling, $A_e$.
The analyzing power of the tags for true $s\bar{s}$ events, as well as the
relative contribution of $u \bar u + d \bar d$ events, were determined 
from the data as described in section 5. This procedure removes much of 
the model dependence.

\section{Apparatus and Hadronic Event Selection}

A general description of the SLD can be found elsewhere~\cite{sld}.
The trigger and initial selection criteria for hadronic $Z^0$ decays are
described in Ref.~\cite{sldalphas}.
This analysis used charged tracks measured in the Central Drift
Chamber (CDC)~\cite{cdc} and Vertex Detector (VXD)~\cite{vxd}, and identified
using the Cherenkov Ring Imaging Detector (CRID)~\cite{crid}.
Momentum measurement is provided by a uniform axial magnetic field of 0.6T.
The CDC and VXD give a momentum resolution of
$\sigma_{p_{\perp}}/p_{\perp}$ = $0.01 \oplus 0.0026p_{\perp}$,
where $p_{\perp}$ is the track momentum transverse to the beam axis in
GeV/$c$.
About 27\% of the data were taken with the original vertex 
detector (VXD2), and the remaining data with the upgraded vertex detector (VXD3).

In the plane normal to the beamline the centroid of the micron-sized SLC 
interaction point (IP) was reconstructed from tracks in sets of approximately 
thirty sequential hadronic $Z^0$ decays to a precision of 
$\sigma_{IP}\simeq 7$ $\mu$m for the VXD2 data and $\simeq 4$ $\mu$m for the
VXD3 data.
Including the uncertainty on the IP position, the resolution on the
charged track impact parameter ($d$) projected in the plane perpendicular
to the beamline is
$\sigma_d =$11$\oplus$70/$(p \sin^{3/2}\theta)$ $\mu$m for VXD2 and
$\sigma_d =$8$\oplus$29/$(p \sin^{3/2}\theta)$ $\mu$m for VXD3, where
$\theta$ is the track polar angle with respect to the beamline.
The CRID comprises two radiator systems which 
identify charged pions
with high efficiency and purity in the momentum range 0.3--35 GeV/c, charged
kaons in the ranges 0.75--6 GeV/c and 9--35 GeV/c, and protons in the ranges
0.75--6 GeV/c and 10--46 GeV/c ~\cite{pprod}.
The event thrust axis~\cite{thrust} was calculated using energy clusters
measured in the Liquid Argon Calorimeter~\cite{lac}.

A set of cuts was applied to the data to select well-measured tracks
and events well contained within the detector acceptance.
Events were required to have the VXD and the CDC operational, 
a minimum of 3 charged tracks with at least
2 VXD hits each, at least 7 charged tracks with $p_{\perp} > 0.2$ GeV/c and
a distance of closest approach within 5 cm along the axis 
from the measured IP, a thrust axis polar angle w.r.t. the beamline, 
$\theta_T$, within $|\cos\theta_T|<0.71$, and a charged visible energy,
$E_{vis}$, of at least 18~GeV, which was calculated from all charged tracks 
by assigning each the charged pion mass.
The efficiency for selecting a well-contained $Z^0 \rightarrow q{\bar q}(g)$
event was estimated to be above 96\% independent of quark flavor.

In order to reduce the effects of decays of heavy hadrons, we selected
light flavor events ($u\bar{u}$, $d\bar{d}$ and $s\bar{s}$) by requiring
at most one high-quality~\cite{homer} track with transverse
impact parameter with respect to the IP of more than 2.5 times its estimated
error to be found in each event.
The selected sample comprised 244,385 events, with an estimated
background contribution 
of 14.4\% from $c\bar{c}$ events, 3.5\% from $b\bar{b}$ 
events, and a non-hadronic background contribution of 
$0.10 \pm 0.05\%$, dominated by $Z^0 \rightarrow \tau^+\tau^-$ events.

For the purpose of estimating the efficiency and purity of the event
flavor tagging and the particle identification, we made use of a detailed 
Monte Carlo (MC) simulation of the detector.
The JETSET 7.4~\cite{jetset} event generator was used, with parameter
values tuned to hadronic $e^+e^-$ annihilation data~\cite{tune},
combined with a simulation of $B$-hadron decays
tuned~\cite{sldsim} to $\Upsilon(4S)$ data and a simulation of the SLD
based on GEANT 3.21~\cite{geant}.
Inclusive distributions of single-particle and event-topology observables
in hadronic events were found to be well described by the
simulation~\cite{sldalphas}.

\section{Selection of $s\bar{s}$ Events}

After the event selection described in the previous section, $s\bar{s}$ 
events are selected by the presence of identified high-momentum $K^\pm$, 
$K_s^0$ or $\Lambda^0$/$\bar\Lambda^0$. These particles are 
likely~\cite{lpprl} to contain an initial $s/ \bar s$ quark, but could also 
contain an initial $u$ and/or $d$ quark or be from the decay of a 
D or B hadron. 
In this analysis the strategy for reducing the model dependence of the result 
involves hard analysis cuts to suppress the non-$s\bar s$ background 
and enhance the analyzing power of the signal to a level where useful 
constraints can be obtained from the data. 

\begin{figure}[htb]
 \centering
 \epsfxsize14cm
 \epsfxsize14cm
 \leavevmode
 \epsffile{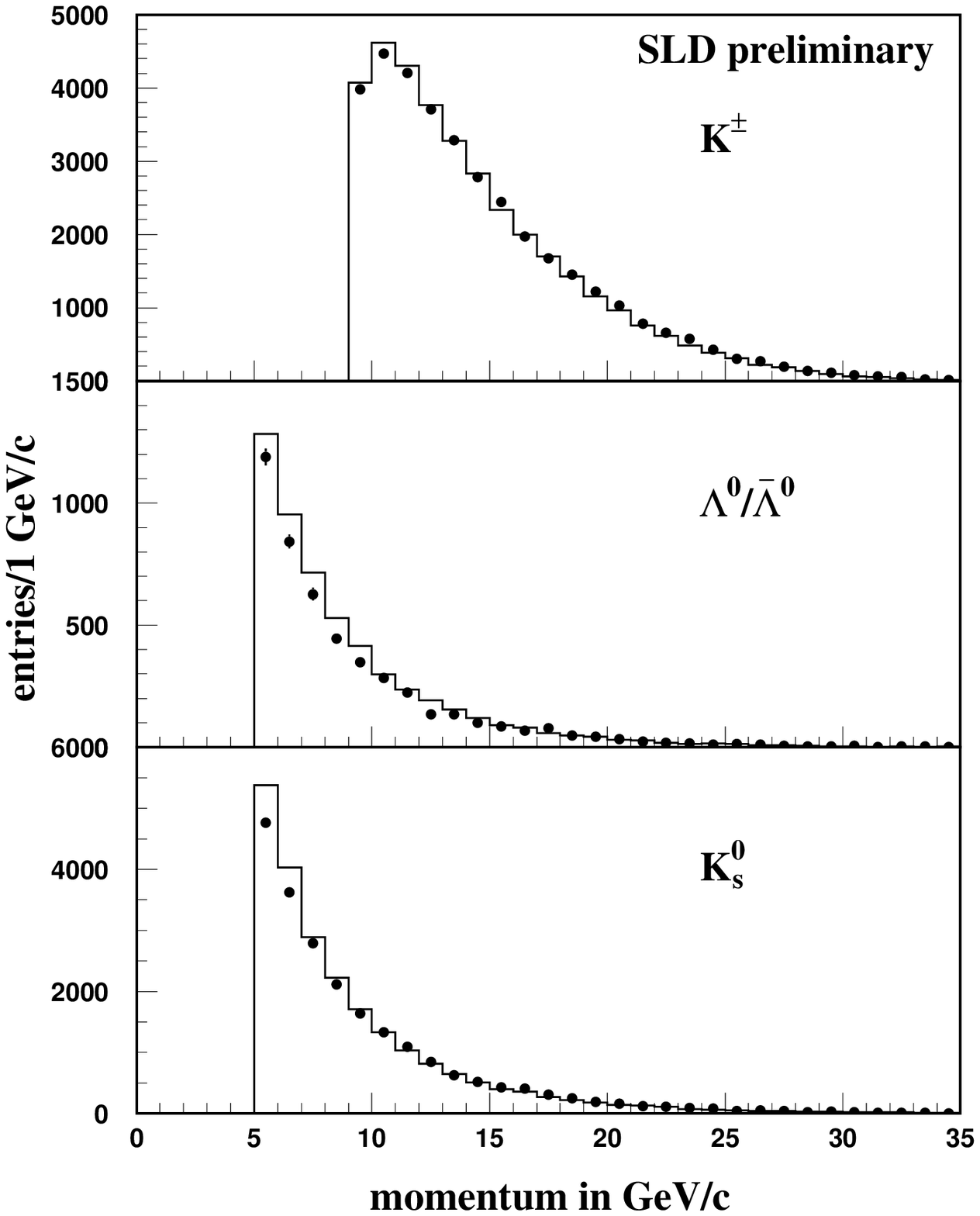}
 \caption{Momentum distributions for selected 
          (a) $K^\pm$, (b) $\Lambda^0/\bar \Lambda^0$ and (c) $K_s^0$ candidates 
          in the data (dots). Also shown is the Monte Carlo simulation (histogram). 
          The Monte Carlo distributions for $\Lambda^0/\bar \Lambda^0$ candidates 
          and $K_s^0$ candidates were later corrected, as described in the text.}
 \label{specdmc}
 \end{figure}

The first step is the selection of strange particles.
The CRID allows $K^\pm$ to be separated from $p /\bar p$ and $\pi^\pm$ with 
high purity for tracks with $p > 9$ GeV/c as described in detail in~\cite{pprod}.
For the purpose of identifying $K^\pm$, relatively loose 
quality cuts are applied. 
Tracks with poor CRID information or that are likely to have scattered or
interacted before exiting the CRID are removed by requiring each track to have
a distance of closest approach transverse to the beam axis within 1 mm,
and within 5 mm along the axis from the measured IP,
to extrapolate through an active region of the CRID gas radiator and 
through a live CRID TPC.
For the remaining tracks log-likelihoods
~\cite{pprod,llik} are calculated for the CRID gas radiator for each 
of the three charged hadron hypotheses $\pi^\pm$, $K^\pm$ and $p/\bar p$.
A track is tagged as a $K^\pm$ by the gas system if the log-likelihood for 
this hypothesis exceeds both of the other
log-likelihoods by at least 3 units. 
Figure~\ref{specdmc} shows the momentum distribution of identified $K^\pm$ for 
our data and Monte Carlo simulation. 
The data and simulation are in quite good agreement. The average purity of 
the $K^\pm$ sample was estimated using the simulation to be 91.5\%. 

The selection of $K_s^0$ and $\Lambda^0$/$\bar\Lambda^0$ is also described
in detail in~\cite{pprod}. 
Briefly, $K_s^0$ and $\Lambda^0$/$\bar\Lambda^0$ are reconstructed in the 
modes $K_s^0 \rightarrow \pi^+ \pi^-$ (BR $\approx$ 69\%) and 
$\Lambda^0 (\bar \Lambda^0) \rightarrow p (\bar p) \pi^\mp$ (BR $\approx$ 64\%) 
and are identified by their long flight distance, reconstructed mass, and 
accuracy of pointing back to the primary interaction point.

For the selection of $K_s^0$ and $\Lambda^0$/$\bar\Lambda^0$, we required
a track acceptance of $|\cos\theta|<0.9$ and at least 30 CDC hits 
for each track.
$K_s^0$ and $\Lambda^0$/$\bar\Lambda^0$ are required to have $p > 5$ GeV/c
and a flight distance with respect to the IP of more than 5 times their 
estimated uncertainties. Gamma conversions are removed by requiring
$m_{ee} > 100$ MeV/$c^2$. 
The $K_s^0$ and $\Lambda^0$/$\bar\Lambda^0$ mass cuts mentioned below are 
parametrized as a function of momentum to take into account the 
dependence of the $m_{p\pi}$ and $m_{\pi\pi}$ mass resolution, $\sigma$, 
on momentum.

In the case of the $\Lambda^0$/$\bar\Lambda^0$, we next use information 
from the Cherenkov Ring Imaging Detector to identify the $p/ \bar p$ 
candidate if it passes above cuts.
We identify the $p/ \bar p$ 
candidate if the log-likelihood for this hypothesis exceeds the 
log-likelihood for the $\pi^+/\pi^-$ hypothesis.
If CRID information on the $p/ \bar p$ candidate 
is not available, we increase the
cut on the flight distance with respect to the IP, normalized by its
estimated uncertainty, to 10,
and require the $m_{\pi\pi}$ of the candidate not to be within $2 \sigma$ of
the nominal $K_s^0$ mass.
Finally, $\Lambda^0$/$\bar\Lambda^0$ are identified by requiring the invariant 
mass of pairs of tracks, $m_{p \pi}$,
to be within $2 \sigma$ of the nominal $\Lambda^0$/$\bar\Lambda^0$ mass.
Figure~\ref{specdmc} gives the momentum distribution for the total selected
$\Lambda^0$/$\bar\Lambda^0$ sample.
The Monte Carlo simulation predicts too many low-momentum 
$\Lambda^0$/$\bar\Lambda^0$ candidates.
We correct this discrepancy in the simulation
by applying a momentum-independent correction factor to the number 
of simulated true $\Lambda^0$/$\bar\Lambda^0$ candidates 
with $p < 15$ GeV/c; this procedure rejects a total of 12.5\% of the 
simulated true $\Lambda^0$/$\bar\Lambda^0$ sample within this momentum region
but keeps the absolute background level which is seen from $m_{p \pi}$
sidebands to be well simulated.
The effect of this correction on the final result will be discussed in
section 5. The corrected simulation predicts that the purity of the 
$\Lambda^0$/$\bar\Lambda^0$ sample is 90.7\%.

Pairs of tracks with invariant mass $m_{\pi\pi}$ within $2 \sigma$ of 
the nominal $K_s^0$ mass are identified as $K_s^0$. 
Figure~\ref{specdmc} gives the momentum distribution for the $K_s^0$ sample.
The Monte Carlo simulation for the $K_s^0$ momentum has an excess for
low momenta, and similar to the case in the $\Lambda^0$/$\bar\Lambda^0$ sample,
we correct this discrepancy in the simulation
by applying a momentum-independent correction factor to the number 
of simulated true $K_s^0$ candidates with $p < 10$ GeV/c; this 
procedure rejects a total of 6.9\% of the simulated true $K_s^0$ 
sample within this momentum region but keeps the absolute background 
level which is seen from $m_{\pi \pi}$ sidebands to be well simulated.
The effect of this correction on the final result will be discussed in
section 5. The corrected simulation predicts that the purity 
of the $K_s^0$ sample is 90.6\%.

These strange particles are then used to tag $s$ and $\bar s$ jets as follows.
Each event is divided into two hemispheres by a plane perpendicular to the 
thrust axis.
We require each of the two hemispheres to contain at least one identified 
strange particle ($K^\pm$, $K_s^0$ or $\Lambda^0 / \bar \Lambda^0$); for 
hemispheres with multiple strange particles we only consider the one with 
the highest momentum.
We require at least one of the two hemispheres to have definite strangeness 
(i.e. to contain a $K^\pm$ or $\Lambda^0 / \bar \Lambda^0$). In events with two 
hemispheres of definite strangeness, the two hemispheres 
are required to have opposite strangeness (e.g. $K^+ K^-$).
This procedure increases the $s\bar{s}$ purity substantially compared with a 
single tag; thus, for these events, the model dependence of the measurement 
(section 4) is reduced.
Table~\ref{dtags} summarizes the composition of the selected event sample for 
data and simulation for each of the 5 tagging modes used. 
The number of events for each mode shown is in good agreement with the 
Monte Carlo prediction. The $s \bar s$ purity and $s \bar s$ analyzing power 
were estimated from the data as discussed below.

\begin{table}[hbt]
\begin{center}
\caption[Summary of selected event sample]{\label{dtags} 
Summary of the selected event sample for 5 modes in data and simulation.}
\bigskip
\begin{tabular}{|c||c|c|c|c|} \hline
Mode & \# Data Events & MC prediction & $s \bar s$ purity & $s \bar s$ analyzing power \\
\hline \hline

$K^+ K^-$                               & 1290 & 1312 & 0.73 & 0.95  \\
\hline
$K^+ \Lambda^0, K^- \bar \Lambda^0$     &  218 &  213 & 0.65 & 0.89  \\
\hline
$\Lambda^0 \bar \Lambda^0$              &   17 &   14 & 0.52 & 0.60  \\
\hline
$K^\pm K_s^0$                           & 1580 & 1614 & 0.61 & 0.70  \\
\hline
$\Lambda^0 K_s^0, \bar \Lambda^0 K_s^0$ &  189 &  194 & 0.50 & 0.35  \\

\hline \hline
Total:                                  & 3294 & 3347 & 0.65 & 0.81  \\

\hline
\end{tabular}
\end{center}
\end{table}

The $K^\pm K^\mp$ mode and the $K^\pm K_s^0$ mode dominate the sample and the 
$K^\pm K^\mp$ mode has the highest $s\bar{s}$ purity. 
The combined $s\bar{s}$ purity of all modes is 65\%, and the predicted 
background in the selected event sample consists 
of 9\% $u \bar u$, 9\% $d \bar d$, 16\% $c \bar c$, and 1\% $b \bar b$ events.
 
The analyzing power is defined as:
    \begin{equation}
      a_s = \frac{N^{right}_s - N^{wrong}_s}{N^{right}_s + N^{wrong}_s}  
\end{equation}
where $N^{right}_s (N^{wrong}_s)$ denotes the number of $s\bar s$ events
in which a particle of negative strangeness is found in the true $s (\bar s)$ 
hemisphere. The average analyzing power for all modes is predicted by the 
simulation to be 0.81. 
The $K^\pm K^\mp$ mode has a substantially higher analyzing power than the 
other modes.

The initial $s$ quark direction is approximated by the thrust axis, $\hat t$ of 
the event, signed to point in the direction of negative strangeness:
\begin{equation}
 x = cos \theta_s = S  \frac{\vec{p} \cdot \hat{t}}{|\vec{p} \cdot \hat{t}|}  
     \hat t_z,
\end{equation}
where $S$ and $\vec{p}$ denote the strangeness and the momentum of the tagging 
particle. 

\begin{figure}[htb]
 \centering
 \epsfxsize15cm
 \epsfxsize15cm
 \leavevmode
 \epsffile{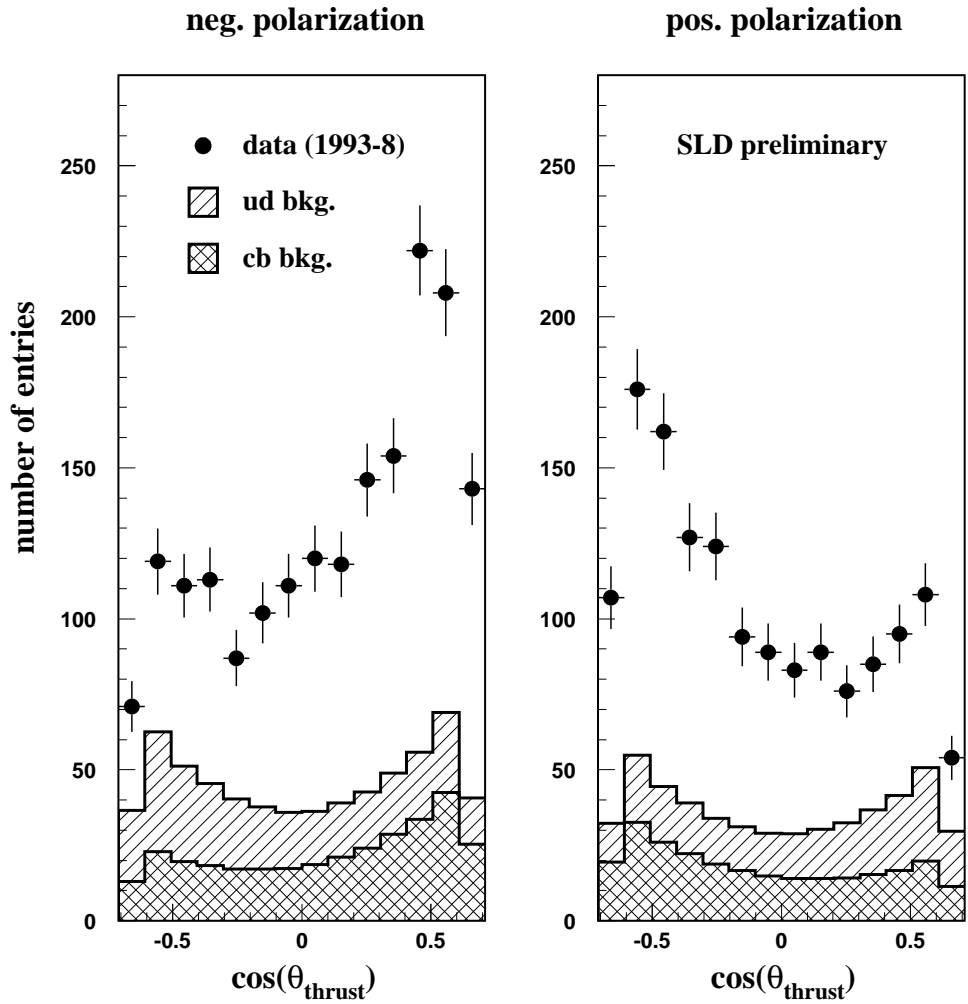}
 \caption{Polar angle distributions of the tagged strange quark, 
          for negative (left) and positive (right) beam polarization. 
          The dots show data, and our estimates of the non-$s \bar s$ backgrounds 
          are indicated by the hatched histograms.}
 \label{asymplot}
 \end{figure}

Figure~\ref{asymplot} shows the polar angle distributions, for all modes 
combined, of the tagged strange quark, for left-handed and right-handed 
electron beams. The expected production asymmetries, of opposite sign for 
the left-handed and the right-handed beams, are clearly visible.

\section{Extraction of $A_s$}

$A_s$ is extracted from these distributions by an unbinned maximum 
likelihood fit. The likelihood function is given by:

\begin{equation}
    L = \prod_{k = 1}^{N_{data}} 
          \{(1 - A_e P_e)(1 + x^2_k) + 
            2 (A_e - P_e) \sum_{f}^{} (N_f [1 + \delta] a_f A_f x_k)\}. 
\end{equation} 

\noindent
Here, $N_f = N_{events} R_f \epsilon_f$ denotes the number of events in the sample 
of flavor $f$ ($f = u,d,s,c,b$) in terms of the number of selected 
hadronic events $N_{events}$, 
$R_f = \Gamma(Z^0 \rightarrow f \bar f)/\Gamma(Z^0 \rightarrow$ hadrons$)$
and the tagging efficiencies $\epsilon_f$; 
$\delta = -0.013$ corrects for the effects of hard gluon radiation~\cite{stavolsen};
$a_f$ denotes the analyzing power for tagging the $f$ rather than the 
$\bar f$ direction; and
$A_f$ is the coupling parameter for flavor $f$.
The parameters $\epsilon_c$, $\epsilon_b$, and $a_c$, $a_b$ for the heavy flavors are 
taken from the Monte Carlo simulation~\cite{sldsim} since a number of independent 
measurements lead us to believe these parameters to be reliable within well defined 
uncertainties. The world average experimental measurements of the parameters $A_c$, 
$A_b$, $R_c$, $R_b$~\cite{ewlepslc} were used. The corresponding systematic 
uncertainties are small and are discussed below.

For the light flavors, the relevant parameters in the fitting 
function are derived where possible from the data. The total number of light 
flavor events, $N_{uds}$, 
is determined by subtracting the number of heavy flavor events (obtained from the 
simulation) from the entire event sample. 
The values for the ratio $N_{ud}/N_s$ and the $s\bar s$ analyzing power, $a_s$,
depend on the tagging mode as shown in Table~\ref{dtags}.
As discussed in the next section, the ratio $N_{ud}/N_s$ and 
the $s\bar s$ analyzing power, $a_s$, for each mode are determined from 
the simulation and are constrained using the data.
The $(u\bar u + d\bar d)$ analyzing power, $a_{ud}$, for each mode is 
estimated to be $\pm a_s/2$ (minus sign for $K^+ K^-$, $K^\pm K_s^0$, and
$K^+ \Lambda^0/K^- \bar \Lambda^0$ modes; plus sign for 
$\Lambda^0 K_s^0/ \bar \Lambda^0 K_s^0$ and $\Lambda^0 \bar \Lambda^0$ modes; 
section 5). The coupling parameters $A_u$ and $A_d$ are set to 
the Standard Model values.

The fit quality of the unbinned maximum likelihood fit to the polar angle
distributions, shown in Figure~\ref{asymplot}, is good with a 
$\chi^2$ of 25.0 for 28 bins. Also included are our estimates of 
non-$s \bar s$ background. The cross-hatched histograms indicate 
$c\bar c + b \bar b$ backgrounds which are seen to show asymmetries of the 
same sign and similar magnitude to the total distribution.
The hatched histograms indicate $u\bar u + d \bar d$ backgrounds showing 
asymmetries of the opposite sign and magnitude to the total distribution. 
The $A_s$ value extracted from the fit is $A_s = 0.85 \pm 0.06(stat.)$.

\section{Systematic Uncertainties and Checks}

The understanding of the parameters used as inputs to the fitting function and
of their uncertainties is crucial to this analysis.
The characteristics of heavy flavor events relevant to this analysis have been
measured experimentally, and our simulation~\cite{jetset,tune,geant} has been 
tuned~\cite{sldsim} to reproduce these results.
The effect of uncertainties in the values of $R_c$, $R_b$, $A_c$ and $A_b$ were 
evaluated by varying those parameters by the uncertainties on their world average 
values~\cite{ewlepslc}.
Uncertainties in other measured quantities such as
the $D$ and $B$ hadron fragmentation functions, the number of
$K^-$ and $K^+$ mesons produced per $D$ or $B$ hadron decay, as well as a number of 
other quantities~\cite{sldalphas} were taken into account by varying 
each quantity in turn by plus and minus the error on its world average value.
In each case the simulated events were weighted to approximate a distribution
generated with the parameter value in question, the Monte Carlo predictions for $N_{cb}$
and $a_{cb}$ rederived, a new fit performed, and the difference between the $A_s$
value extracted and the central value taken as a systematic error.

The sum in quadrature of these uncertainties was taken as the systematic
error due to heavy flavor modelling and is listed in Table~2.
This is a relatively small contribution to the total systematic error.
Other small contributions to the systematic error include those from the 0.6\% uncertainty 
in the correction for the effect of hard gluon radiation, and the 0.8\% uncertainty 
in the beam polarization. 

For the light flavors, there are few experimental constraints on the relevant 
input parameters.
Qualitative features such as leading particle production~\cite{lpprl}, short 
range rapidity correlations between high-momentum $K K$ and baryon-antibaryon 
pairs~\cite{davecorrel} and long-range correlations between several particle 
species~\cite{davecorrel} have been observed experimentally, but these results are not 
sufficient to quantify the analyzing power of the strange-particle tag or the $u \bar u$ 
and $d \bar d$ background. Our Monte Carlo simulation provides a reasonable 
description of the above observations, and we have used our data to constrain 
the relevant input parameters in the context of our Monte Carlo model.

For the analyzing power in $s \bar{s}$ events, we note that there are only two ways 
to mis-tag an $s$ jet as an $\bar{s}$ jet: either the jet must contain a true $K^+$ 
or $\bar{\Lambda}^0$ that satisfies our cuts, or we must mis-identify a $\pi^+$ or $p$ 
as a $K^+$ or reconstruct a fake $\bar{\Lambda}^0$.
The Monte Carlo simulation predicts that the fraction of events with a mis-identified 
particle is negligible in tagged $s \bar{s}$ events, since the majority of high-momentum 
tracks in these events are kaons, and the relative $V^0$ fake rate is low.
We have measured our mis-identification rates in the data~\cite{pprod}, and they 
contribute less than 0.2\% to the wrong sign fraction, so we neglect this source of 
systematic uncertainty.

If a non-leading high-momentum $K^+$ is produced in an $s$ jet, then there must be an
associated strange particle in the jet, which will also tend to have high
momentum.
Including the leading strange particle, such a jet will contain one
antistrange and two strange particles, all with relatively high momentum.
We can therefore investigate the rate of production of these wrong-sign kaons by 
studying events in which we find three identified $K^{\pm}$ and/or $K_s^0$ in the 
same hemisphere.
Such an event sample is expected to be fairly pure in $s$/$\bar{s}$, since a $u$/$\bar{u}$ 
or $d$/$\bar{d}$ jet would have to contain either four strange particles or two strange 
particles and one mis-identified particle in order to be selected.
In our data we found 68 hemispheres containing three identified 
$K^{\pm}$ and/or $K_s^0$, compared with a Monte Carlo prediction of 73.
We subtracted the simulated non-s jet background of 27 events
to yield a measured number of $41 \pm 9$ jets with 3 kaons, providing a 20\% constraint 
on the number of $s \bar s$ events that {\it could} have the wrong sign.
Since the Monte Carlo prediction is consistent with the data, we used the simulated 
$a_s$ for each mode (see Table~\ref{dtags}) as our central value for the analyzing 
power in $s\bar{s}$ events.
This constraint is not entirely model-independent, since we are relying on the model to 
predict the fraction of these jets in which all three kaons pass our momentum cuts, as 
well as the fraction in which the wrong-sign kaon is chosen as the tagging particle rather 
than either of the right-sign kaons.
We also assume equal production of charged versus neutral kaons (as in the Monte Carlo
simulation); thus, this procedure delivers a simultaneous calibration of the
analyzing power in $s \bar{s}$ events for the $K^+ K^-$ and $K^{\pm} K_s^0$ modes.
However, we trust the Monte Carlo simulation for the modes involving 
$\Lambda^0$/$\bar\Lambda^0$.
Therefore, we conservatively applied the 20\% uncertainty to the wrong-sign 
fraction of each tagging mode, resulting in a 3\% uncertainty on $A_s$, as shown 
in Table~\ref{systerr}. We also counted hemispheres containing 
a $K^+ K^+$ or $K^- K^-$ pair, obtaining a consistent but less precise constraint.

\begin{table}[hbt]
\begin{center}
\caption[Summary systematic uncertainties]{\label{systerr} Summary of systematic uncertainties.}
\bigskip
\begin{tabular}{|c||c|c|c|} \hline
Source & Comments & Systematic variation & $\delta A_s / A_s$ \\
\hline \hline
           
heavy flavor modelling  & MC/world averages    & Ref.~\cite{ewlepslc,tune,sldsim} & $0.012$ \\
\hline 
hard gluon radiation    & Stav-Olsen with      & $(1.3 \pm 0.6)\%$   & $0.006$ \\
                        & bias correction      &                     &         \\
\hline 
beam polarization       & data                 & $(73.4 \pm 0.8)\%$  & $0.011$ \\
\hline 
$a_s$                   & MC constrained by    & $\pm 20\%$ on       & $0.034$ \\
                        & 3 K jets in data     & wrong sign fraction &         \\
\hline
$a_{ud}$                & $a_{ud} = \pm a_s/2$ & $\pm 57.7\%$        & $0.057$ \\
$A_{ud}$                & Standard Model       &  --                 & --      \\
\hline
$N_{ud} / N_s$          & MC constrained by    & $\pm 8.4\%$         & $0.028$ \\
                        & 2 K jets in data     &                     &         \\
\hline 
MC statistics           &                      &                     & 0.019   \\         
\hline \hline
Total:                  &                      &                     & $0.077$ \\

\hline 
\end{tabular}
\end{center}
\end{table}

The relative $u \bar u + d \bar d$ background level $N_{ud}/N_s$ was 
constrained from the data by exploiting the fact that an even number of strange 
particles must be produced in a u/d jet, and that they appear in strange-antistrange 
pairs that have 
similar momenta. We counted 1262 hemispheres in the data containing an identified 
$K^+$-$K^-$ pair and 983 hemispheres containing an identified $K^\pm$-$K^0$ pair.
The respective Monte Carlo predictions of 1215 and 1005 are consistent.
After subtracting the predicted non-$u$/$d$ jet backgrounds, these two checks 
yielded 9\% and 8\% constraints, respectively, on the $u \bar u + d \bar d$ background.
We also counted events in the data that were tagged by kaons of the same sign in 
both hemispheres. The Monte Carlo prediction is consistent, but the constraint obtained
is less precise.
Again, we have used the Monte Carlo central value for each tagging mode and, since the 
constraints are not completely model-independent, we have used only the most precise one to 
estimate the systematic uncertainty.

The above checks are also sensitive to the analyzing power of 
$u\bar u + d\bar d$ events, $a_{ud}$.
However, with the present event statistics we cannot obtain a tight 
constraint on this quantity.
We therefore assume that $a_{ud}$ must be negative (positive) for the 
$K^+ K^-$, $K^\pm K_s^0$, and $K^+ \Lambda^0/K^- \bar \Lambda^0$ modes   
($\Lambda^0 K_s^0/ \bar \Lambda^0 K_s^0$ and $\Lambda^0 \bar \Lambda^0$ modes), 
since $u$ and $d$ jets must produce 
a leading $K^+$ ($\Lambda^0$) rather than $K^-$ ($\bar \Lambda^0$), 
and that the modulus of $a_{ud}$ must be less 
than that of $a_s$, since there is always a companion particle of opposite 
strangeness in a $u$ or $d$ jet that will tend to dilute the analyzing power. 
For all 5 tagging modes, we take these as hard limits, 
$0 < |a_{ud}| < a_s$, use the middle of the range for our central value 
and assign an uncertainty equal to the range divided by $\sqrt{12}$. 
The shift in central value of $A_s$ due to this estimation of $a_{ud}$,
as compared to the simulated values for $a_{ud}$, was found to be negligible.

The effects on the central value of $A_s$ due to the corrections (section 3) of the 
Monte Carlo $K_s^0$ and $\Lambda^0$/$\bar\Lambda^0$ momentum distributions 
were studied. It was determined that the changes in the $s \bar s$ purity and the 
analyzing power in $s \bar s$ events were small.
The change in the central value of $A_s$ when these corrections were removed 
was smaller than any of the contributions listed in Table~2 and we considered it 
to be negligible. The individual systematic errors were added in quadrature to yield a total
systematic error of $\delta A_s / A_s = 0.077$, i.e. $\delta A_s = 0.07$.

\section{Summary and Conclusion}

We have presented a preliminary direct measurement of the parity violating coupling 
of the $Z^0$ to strange quarks, $A_s$, derived from the sample of approximately 550,000 
hadronic decays of Z$^{0}$ bosons produced with a polarized electron beam and 
recorded by the SLD experiment at SLAC between 1993 and 1998.
The precision CCD vertex detector allows the suppression of the heavy flavor 
background, and the Cherenkov Ring Imaging Detector is crucial in the tagging
of high-momentum $K^\pm$ and helps improve the $\Lambda^0$/$\bar\Lambda^0$ purity.
The coupling $A_s$ is obtained directly from a measurement of the 
left-right-forward-backward production asymmetry in polar angle of the tagged 
$s$ quark.
The background from $u\bar{u}$ and $d\bar{d}$ events is measured from the 
data, as is the analyzing power of the method for $s\bar{s}$ events.

An unbinned maximum likelihood fit is used to obtain the result:  
\begin{equation}
    A_s  = 0.85 \pm 0.06(stat.) \pm 0.07 (syst.) (preliminary).
\end{equation}    

\noindent
This result is consistent with the Standard Model expectation for $A_s$.
Our measurement can be used to test the universality of the coupling constants
by comparing it with the world average value for $A_b$~\cite{ewlepslc}. 
The two measurements are consistent.

In order to compare with previous measurements of $A_{FB}^{s}$ and $A_{FB}^{d,s}$
(see section 1), we must assume a value of $A_e$. Using $A_e = 0.1499$~\cite{ewlepslc}
and neglecting the small uncertainty on $A_e$, 
the DELPHI measurements translate into
$A_s = 1.165 \pm 0.311(stat.) \pm 0.116(syst.)$
and
$A_{d,s} = 0.996 \pm 0.276(stat.) \pm 0.480(syst.)$. 
Similarly, the OPAL measurement yields
$A_{d,s} = 0.605 \pm 0.311(stat.) \pm 0.098(syst.)$.
Our measurement is consistent with these and represents a substantial
improvement in precision.

\vskip .5truecm

\section*{Acknowledgements}
We thank the personnel of the SLAC accelerator department and the
technical
staffs of our collaborating institutions for their outstanding efforts
on our behalf.

\vskip 1truecm

\vbox{
\footnotesize\renewcommand{\baselinestretch}{1}
\noindent
 This work was supported by Department of Energy
  contracts:
  DE-FG02-91ER40676 (BU),
  DE-FG03-91ER40618 (UCSB),
  DE-FG03-92ER40689 (UCSC),
  DE-FG03-93ER40788 (CSU),
  DE-FG02-91ER40672 (Colorado),
  DE-FG02-91ER40677 (Illinois),
  DE-AC03-76SF00098 (LBL),
  DE-FG02-92ER40715 (Massachusetts),
  DE-FC02-94ER40818 (MIT),
  DE-FG03-96ER40969 (Oregon),
  DE-AC03-76SF00515 (SLAC),
  DE-FG05-91ER40627 (Tennessee),
  DE-FG02-95ER40896 (Wisconsin),
  DE-FG02-92ER40704 (Yale);
  National Science Foundation grants:
  PHY-91-13428 (UCSC),
  PHY-89-21320 (Columbia),
  PHY-92-04239 (Cincinnati),
  PHY-95-10439 (Rutgers),
  PHY-88-19316 (Vanderbilt),
  PHY-92-03212 (Washington);
  the UK Particle Physics and Astronomy Research Council
  (Brunel, Oxford and RAL);
  the Istituto Nazionale di Fisica Nucleare of Italy
  (Bologna, Ferrara, Frascati, Pisa, Padova, Perugia);
  the Japan-US Cooperative Research Project on High Energy Physics
  (Nagoya, Tohoku);
  and the Korea Science and Engineering Foundation (Soongsil).}
  

\vfill
\eject

\section*{$^{**}$List of Authors}
%
%
%
\begin{center}
\def\iADEL{$^{(1)}$}
\def\iAOMORI{$^{(2)}$}
\def\iBOLO{$^{(3)}$}
\def\iBRI{$^{(4)}$}
\def\iBRUN{$^{(5)}$}
\def\iBU{$^{(6)}$}
\def\iCINC{$^{(7)}$}
\def\iCOLO{$^{(8)}$}
\def\iCOLU{$^{(9)}$}
\def\iCSU{$^{(10)}$}
\def\iFERR{$^{(11)}$}
\def\iFRAS{$^{(12)}$}
\def\iILLI{$^{(13)}$}
\def\iJHU{$^{(14)}$}
\def\iLBL{$^{(15)}$}
\def\iLTU{$^{(16)}$}
\def\iMASS{$^{(17)}$}
\def\iMISSI{$^{(18)}$}
\def\iMIT{$^{(19)}$}
\def\iMOSCOW{$^{(20)}$}
\def\iNAGO{$^{(21)}$}
\def\iOREG{$^{(22)}$}
\def\iOXF{$^{(23)}$}
\def\iPADO{$^{(24)}$}
\def\iPERU{$^{(25)}$}
\def\iPISA{$^{(26)}$}
\def\iRAL{$^{(27)}$}
\def\iRUTG{$^{(28)}$}
\def\iSLAC{$^{(29)}$}
\def\iSOGA{$^{(30)}$}
\def\iSOONG{$^{(31)}$}
\def\iTENN{$^{(32)}$}
\def\iTOHO{$^{(33)}$}
\def\iUCSB{$^{(34)}$}
\def\iUCSC{$^{(35)}$}
\def\iUVIC{$^{(36)}$}
\def\iVAND{$^{(37)}$}
\def\iWASH{$^{(38)}$}
\def\iWISC{$^{(39)}$}
\def\iYALE{$^{(40)}$}

  \baselineskip=.75\baselineskip  
\mbox{Kenji  Abe\unskip,\iNAGO}
\mbox{Koya Abe\unskip,\iTOHO}
\mbox{T. Abe\unskip,\iSLAC}
\mbox{I.Adam\unskip,\iSLAC}
\mbox{T.  Akagi\unskip,\iSLAC}
\mbox{N. J. Allen\unskip,\iBRUN}
\mbox{W.W. Ash\unskip,\iSLAC}
\mbox{D. Aston\unskip,\iSLAC}
\mbox{K.G. Baird\unskip,\iMASS}
\mbox{C. Baltay\unskip,\iYALE}
\mbox{H.R. Band\unskip,\iWISC}
\mbox{M.B. Barakat\unskip,\iLTU}
\mbox{O. Bardon\unskip,\iMIT}
\mbox{T.L. Barklow\unskip,\iSLAC}
\mbox{G. L. Bashindzhagyan\unskip,\iMOSCOW}
\mbox{J.M. Bauer\unskip,\iMISSI}
\mbox{G. Bellodi\unskip,\iOXF}
\mbox{R. Ben-David\unskip,\iYALE}
\mbox{A.C. Benvenuti\unskip,\iBOLO}
\mbox{G.M. Bilei\unskip,\iPERU}
\mbox{D. Bisello\unskip,\iPADO}
\mbox{G. Blaylock\unskip,\iMASS}
\mbox{J.R. Bogart\unskip,\iSLAC}
\mbox{G.R. Bower\unskip,\iSLAC}
\mbox{J. E. Brau\unskip,\iOREG}
\mbox{M. Breidenbach\unskip,\iSLAC}
\mbox{W.M. Bugg\unskip,\iTENN}
\mbox{D. Burke\unskip,\iSLAC}
\mbox{T.H. Burnett\unskip,\iWASH}
\mbox{P.N. Burrows\unskip,\iOXF}
\mbox{A. Calcaterra\unskip,\iFRAS}
\mbox{D. Calloway\unskip,\iSLAC}
\mbox{B. Camanzi\unskip,\iFERR}
\mbox{M. Carpinelli\unskip,\iPISA}
\mbox{R. Cassell\unskip,\iSLAC}
\mbox{R. Castaldi\unskip,\iPISA}
\mbox{A. Castro\unskip,\iPADO}
\mbox{M. Cavalli-Sforza\unskip,\iUCSC}
\mbox{A. Chou\unskip,\iSLAC}
\mbox{E. Church\unskip,\iWASH}
\mbox{H.O. Cohn\unskip,\iTENN}
\mbox{J.A. Coller\unskip,\iBU}
\mbox{M.R. Convery\unskip,\iSLAC}
\mbox{V. Cook\unskip,\iWASH}
\mbox{R. Cotton\unskip,\iBRUN}
\mbox{R.F. Cowan\unskip,\iMIT}
\mbox{D.G. Coyne\unskip,\iUCSC}
\mbox{G. Crawford\unskip,\iSLAC}
\mbox{C.J.S. Damerell\unskip,\iRAL}
\mbox{M. N. Danielson\unskip,\iCOLO}
\mbox{M. Daoudi\unskip,\iSLAC}
\mbox{N. de Groot\unskip,\iBRI}
\mbox{R. Dell'Orso\unskip,\iPERU}
\mbox{P.J. Dervan\unskip,\iBRUN}
\mbox{R. de Sangro\unskip,\iFRAS}
\mbox{M. Dima\unskip,\iCSU}
\mbox{A. D'Oliveira\unskip,\iCINC}
\mbox{D.N. Dong\unskip,\iMIT}
\mbox{M. Doser\unskip,\iSLAC}
\mbox{R. Dubois\unskip,\iSLAC}
\mbox{B.I. Eisenstein\unskip,\iILLI}
\mbox{V. Eschenburg\unskip,\iMISSI}
\mbox{E. Etzion\unskip,\iWISC}
\mbox{S. Fahey\unskip,\iCOLO}
\mbox{D. Falciai\unskip,\iFRAS}
\mbox{C. Fan\unskip,\iCOLO}
\mbox{J.P. Fernandez\unskip,\iUCSC}
\mbox{M.J. Fero\unskip,\iMIT}
\mbox{K.Flood\unskip,\iMASS}
\mbox{R. Frey\unskip,\iOREG}
\mbox{J. Gifford\unskip,\iUVIC}
\mbox{T. Gillman\unskip,\iRAL}
\mbox{G. Gladding\unskip,\iILLI}
\mbox{S. Gonzalez\unskip,\iMIT}
\mbox{E. R. Goodman\unskip,\iCOLO}
\mbox{E.L. Hart\unskip,\iTENN}
\mbox{J.L. Harton\unskip,\iCSU}
\mbox{A. Hasan\unskip,\iBRUN}
\mbox{K. Hasuko\unskip,\iTOHO}
\mbox{S. J. Hedges\unskip,\iBU}
\mbox{S.S. Hertzbach\unskip,\iMASS}
\mbox{M.D. Hildreth\unskip,\iSLAC}
\mbox{J. Huber\unskip,\iOREG}
\mbox{M.E. Huffer\unskip,\iSLAC}
\mbox{E.W. Hughes\unskip,\iSLAC}
\mbox{X.Huynh\unskip,\iSLAC}
\mbox{H. Hwang\unskip,\iOREG}
\mbox{M. Iwasaki\unskip,\iOREG}
\mbox{D. J. Jackson\unskip,\iRAL}
\mbox{P. Jacques\unskip,\iRUTG}
\mbox{J.A. Jaros\unskip,\iSLAC}
\mbox{Z.Y. Jiang\unskip,\iSLAC}
\mbox{A.S. Johnson\unskip,\iSLAC}
\mbox{J.R. Johnson\unskip,\iWISC}
\mbox{R.A. Johnson\unskip,\iCINC}
\mbox{T. Junk\unskip,\iSLAC}
\mbox{R. Kajikawa\unskip,\iNAGO}
\mbox{M. Kalelkar\unskip,\iRUTG}
\mbox{Y. Kamyshkov\unskip,\iTENN}
\mbox{H.J. Kang\unskip,\iRUTG}
\mbox{I. Karliner\unskip,\iILLI}
\mbox{H. Kawahara\unskip,\iSLAC}
\mbox{Y. D. Kim\unskip,\iSOGA}
\mbox{M.E. King\unskip,\iSLAC}
\mbox{R. King\unskip,\iSLAC}
\mbox{R.R. Kofler\unskip,\iMASS}
\mbox{N.M. Krishna\unskip,\iCOLO}
\mbox{R.S. Kroeger\unskip,\iMISSI}
\mbox{M. Langston\unskip,\iOREG}
\mbox{A. Lath\unskip,\iMIT}
\mbox{D.W.G. Leith\unskip,\iSLAC}
\mbox{V. Lia\unskip,\iMIT}
\mbox{C.Lin\unskip,\iMASS}
\mbox{M.X. Liu\unskip,\iYALE}
\mbox{X. Liu\unskip,\iUCSC}
\mbox{M. Loreti\unskip,\iPADO}
\mbox{A. Lu\unskip,\iUCSB}
\mbox{H.L. Lynch\unskip,\iSLAC}
\mbox{J. Ma\unskip,\iWASH}
\mbox{G. Mancinelli\unskip,\iRUTG}
\mbox{S. Manly\unskip,\iYALE}
\mbox{G. Mantovani\unskip,\iPERU}
\mbox{T.W. Markiewicz\unskip,\iSLAC}
\mbox{T. Maruyama\unskip,\iSLAC}
\mbox{H. Masuda\unskip,\iSLAC}
\mbox{E. Mazzucato\unskip,\iFERR}
\mbox{A.K. McKemey\unskip,\iBRUN}
\mbox{B.T. Meadows\unskip,\iCINC}
\mbox{G. Menegatti\unskip,\iFERR}
\mbox{R. Messner\unskip,\iSLAC}
\mbox{P.M. Mockett\unskip,\iWASH}
\mbox{K.C. Moffeit\unskip,\iSLAC}
\mbox{T.B. Moore\unskip,\iYALE}
\mbox{M.Morii\unskip,\iSLAC}
\mbox{D. Muller\unskip,\iSLAC}
\mbox{V.Murzin\unskip,\iMOSCOW}
\mbox{T. Nagamine\unskip,\iTOHO}
\mbox{S. Narita\unskip,\iTOHO}
\mbox{U. Nauenberg\unskip,\iCOLO}
\mbox{H. Neal\unskip,\iSLAC}
\mbox{M. Nussbaum\unskip,\iCINC}
\mbox{N.Oishi\unskip,\iNAGO}
\mbox{D. Onoprienko\unskip,\iTENN}
\mbox{L.S. Osborne\unskip,\iMIT}
\mbox{R.S. Panvini\unskip,\iVAND}
\mbox{C. H. Park\unskip,\iSOONG}
\mbox{T.J. Pavel\unskip,\iSLAC}
\mbox{I. Peruzzi\unskip,\iFRAS}
\mbox{M. Piccolo\unskip,\iFRAS}
\mbox{L. Piemontese\unskip,\iFERR}
\mbox{K.T. Pitts\unskip,\iOREG}
\mbox{R.J. Plano\unskip,\iRUTG}
\mbox{R. Prepost\unskip,\iWISC}
\mbox{C.Y. Prescott\unskip,\iSLAC}
\mbox{G.D. Punkar\unskip,\iSLAC}
\mbox{J. Quigley\unskip,\iMIT}
\mbox{B.N. Ratcliff\unskip,\iSLAC}
\mbox{T.W. Reeves\unskip,\iVAND}
\mbox{J. Reidy\unskip,\iMISSI}
\mbox{P.L. Reinertsen\unskip,\iUCSC}
\mbox{P.E. Rensing\unskip,\iSLAC}
\mbox{L.S. Rochester\unskip,\iSLAC}
\mbox{P.C. Rowson\unskip,\iCOLU}
\mbox{J.J. Russell\unskip,\iSLAC}
\mbox{O.H. Saxton\unskip,\iSLAC}
\mbox{T. Schalk\unskip,\iUCSC}
\mbox{R.H. Schindler\unskip,\iSLAC}
\mbox{B.A. Schumm\unskip,\iUCSC}
\mbox{J. Schwiening\unskip,\iSLAC}
\mbox{S. Sen\unskip,\iYALE}
\mbox{V.V. Serbo\unskip,\iSLAC}
\mbox{M.H. Shaevitz\unskip,\iCOLU}
\mbox{J.T. Shank\unskip,\iBU}
\mbox{G. Shapiro\unskip,\iLBL}
\mbox{D.J. Sherden\unskip,\iSLAC}
\mbox{K. D. Shmakov\unskip,\iTENN}
\mbox{C. Simopoulos\unskip,\iSLAC}
\mbox{N.B. Sinev\unskip,\iOREG}
\mbox{S.R. Smith\unskip,\iSLAC}
\mbox{M. B. Smy\unskip,\iCSU}
\mbox{J.A. Snyder\unskip,\iYALE}
\mbox{H. Staengle\unskip,\iCSU}
\mbox{A. Stahl\unskip,\iSLAC}
\mbox{P. Stamer\unskip,\iRUTG}
\mbox{H. Steiner\unskip,\iLBL}
\mbox{R. Steiner\unskip,\iADEL}
\mbox{M.G. Strauss\unskip,\iMASS}
\mbox{D. Su\unskip,\iSLAC}
\mbox{F. Suekane\unskip,\iTOHO}
\mbox{A. Sugiyama\unskip,\iNAGO}
\mbox{S. Suzuki\unskip,\iNAGO}
\mbox{M. Swartz\unskip,\iJHU}
\mbox{A. Szumilo\unskip,\iWASH}
\mbox{T. Takahashi\unskip,\iSLAC}
\mbox{F.E. Taylor\unskip,\iMIT}
\mbox{J. Thom\unskip,\iSLAC}
\mbox{E. Torrence\unskip,\iMIT}
\mbox{N. K. Toumbas\unskip,\iSLAC}
\mbox{T. Usher\unskip,\iSLAC}
\mbox{C. Vannini\unskip,\iPISA}
\mbox{J. Va'vra\unskip,\iSLAC}
\mbox{E. Vella\unskip,\iSLAC}
\mbox{J.P. Venuti\unskip,\iVAND}
\mbox{R. Verdier\unskip,\iMIT}
\mbox{P.G. Verdini\unskip,\iPISA}
\mbox{D. L. Wagner\unskip,\iCOLO}
\mbox{S.R. Wagner\unskip,\iSLAC}
\mbox{A.P. Waite\unskip,\iSLAC}
\mbox{S. Walston\unskip,\iOREG}
\mbox{J.Wang\unskip,\iSLAC}
\mbox{S.J. Watts\unskip,\iBRUN}
\mbox{A.W. Weidemann\unskip,\iTENN}
\mbox{E. R. Weiss\unskip,\iWASH}
\mbox{J.S. Whitaker\unskip,\iBU}
\mbox{S.L. White\unskip,\iTENN}
\mbox{F.J. Wickens\unskip,\iRAL}
\mbox{B. Williams\unskip,\iCOLO}
\mbox{D.C. Williams\unskip,\iMIT}
\mbox{S.H. Williams\unskip,\iSLAC}
\mbox{S. Willocq\unskip,\iMASS}
\mbox{R.J. Wilson\unskip,\iCSU}
\mbox{W.J. Wisniewski\unskip,\iSLAC}
\mbox{J. L. Wittlin\unskip,\iMASS}
\mbox{M. Woods\unskip,\iSLAC}
\mbox{G.B. Word\unskip,\iVAND}
\mbox{T.R. Wright\unskip,\iWISC}
\mbox{J. Wyss\unskip,\iPADO}
\mbox{R.K. Yamamoto\unskip,\iMIT}
\mbox{J.M. Yamartino\unskip,\iMIT}
\mbox{X. Yang\unskip,\iOREG}
\mbox{J. Yashima\unskip,\iTOHO}
\mbox{S.J. Yellin\unskip,\iUCSB}
\mbox{C.C. Young\unskip,\iSLAC}
\mbox{H. Yuta\unskip,\iAOMORI}
\mbox{G. Zapalac\unskip,\iWISC}
\mbox{R.W. Zdarko\unskip,\iSLAC}
\mbox{J. Zhou\unskip.\iOREG}

\it
  \vskip \baselineskip                   
  \vskip \baselineskip        
  \baselineskip=.75\baselineskip   
\iADEL
  Adelphi University, Garden City, New York 11530, \break
\iAOMORI
  Aomori University, Aomori , 030 Japan, \break
\iBOLO
  INFN Sezione di Bologna, I-40126, Bologna Italy, \break
\iBRI
  University of Bristol, Bristol, U.K., \break
\iBRUN
  Brunel University, Uxbridge, Middlesex, UB8 3PH United Kingdom, \break
\iBU
  Boston University, Boston, Massachusetts 02215, \break
\iCINC
  University of Cincinnati, Cincinnati, Ohio 45221, \break
\iCOLO
  University of Colorado, Boulder, Colorado 80309, \break
\iCOLU
  Columbia University, New York, New York 10533, \break
\iCSU
  Colorado State University, Ft. Collins, Colorado 80523, \break
\iFERR
  INFN Sezione di Ferrara and Universita di Ferrara, I-44100 Ferrara, Italy, \break
\iFRAS
  INFN Lab. Nazionali di Frascati, I-00044 Frascati, Italy, \break
\iILLI
  University of Illinois, Urbana, Illinois 61801, \break
\iJHU
  Johns Hopkins University, Baltimore, MD 21218-2686, \break
\iLBL
  Lawrence Berkeley Laboratory, University of California, Berkeley, California 94720, \break
\iLTU
  Louisiana Technical University - Ruston,LA 71272, \break
\iMASS
  University of Massachusetts, Amherst, Massachusetts 01003, \break
\iMISSI
  University of Mississippi, University, Mississippi 38677, \break
\iMIT
  Massachusetts Institute of Technology, Cambridge, Massachusetts 02139, \break
\iMOSCOW
  Institute of Nuclear Physics, Moscow State University, 119899, Moscow Russia, \break
\iNAGO
  Nagoya University, Chikusa-ku, Nagoya 464 Japan, \break
\iOREG
  University of Oregon, Eugene, Oregon 97403, \break
\iOXF
  Oxford University, Oxford, OX1 3RH, United Kingdom, \break
\iPADO
  INFN Sezione di Padova and Universita di Padova I-35100, Padova, Italy, \break
\iPERU
  INFN Sezione di Perugia and Universita di Perugia, I-06100 Perugia, Italy, \break
\iPISA
  INFN Sezione di Pisa and Universita di Pisa, I-56010 Pisa, Italy, \break
\iRAL
  Rutherford Appleton Laboratory, Chilton, Didcot, Oxon OX11 0QX United Kingdom, \break
\iRUTG
  Rutgers University, Piscataway, New Jersey 08855, \break
\iSLAC
  Stanford Linear Accelerator Center, Stanford University, Stanford, California 94309, \break
\iSOGA
  Sogang University, Seoul, Korea, \break
\iSOONG
  Soongsil University, Seoul, Korea 156-743, \break
\iTENN
  University of Tennessee, Knoxville, Tennessee 37996, \break
\iTOHO
  Tohoku University, Sendai 980, Japan, \break
\iUCSB
  University of California at Santa Barbara, Santa Barbara, California 93106, \break
\iUCSC
  University of California at Santa Cruz, Santa Cruz, California 95064, \break
\iUVIC
  University of Victoria, Victoria, B.C., Canada, V8W 3P6, \break
\iVAND
  Vanderbilt University, Nashville,Tennessee 37235, \break
\iWASH
  University of Washington, Seattle, Washington 98105, \break
\iWISC
  University of Wisconsin, Madison,Wisconsin 53706, \break
\iYALE
  Yale University, New Haven, Connecticut 06511. \break

\rm
%

\end{center}


\vfill
\eject

\vskip 1truecm

\vspace{1 mm}

 \vspace{1 mm}
 
\end{document}